# A Data Readout Approach for Physics Experiment*

HUANG Xi-Ru(黄锡汝)[1,2; 1)] CAO Ping(曹平)[1,2; 2)]
GAO Li-Wei(高力为)[1,2] ZHENG Jia-Jun(郑佳俊)[1,2]

[1]State Key Laboratory of Particle Detection and Electronics, University of Science and Technology of China, Hefei 230026, China
[2]Anhui Key Laboratory of Physical Electronics, Department of Modern Physics, University of Science and Technology of China, Hefei 230026, China
[3]Institute of High Energy Physics, Chinese Academy of Sciences, Beijing 100049, China

**Abstract:** With the increasing physical event rate and number of electronic channels, traditional readout scheme meets the challenge of improving readout speed caused by the limited bandwidth of crate backplane. In this paper, a high-speed data readout method based on Ethernet is designed for each module to have capability of transmitting data to DAQ. Features of explicitly parallel data transmitting and distributed network architecture make the readout system has advantage of adapting varying requirements of particle physics experiments. Furthermore, to guarantee the readout performance and flexibility, a standalone embedded CPU system is utilized for network protocol stack processing. To receive customized data format and protocol from front-end electronics, a field programmable gate array (FPGA) is used for logic reconfiguration. To optimize the interface and improve the data swap speed between CPU and FPGA, a sophisticated method based on SRAM is presented in this paper. For the purpose of evaluating this high-speed readout method, a simplified readout module is designed and implemented. Test results show that this module can support up to 70Mbps data throughput from the readout module to DAQ smoothly.

## 1 Introduction[1]

In a typical particle physics experiment, the readout system is generally implemented in a standard crate (e.g. VME). The crate contains a variety of electronic modules, generally including: readout modules used for receiving data from the front-end electronics (FEE) and transmitting event data to the crate controller via the crate backplane bus, and a crate controller used for collecting event data and sending event data through Gigabit networks to the data acquisition (DAQ) system in real time [1]. Fig.1 shows a simplified version of the architecture of a typical readout system.

An "event" describes the result of a single reaction between tow colliding particles in nuclear and particle physics. Among the large amounts of events, only a rare event is associated with the physics objects (electrons, muons, photon, etc.) of interest to physicists. In order to take out most of irrelevant events and reduce the challenges of storage, it is essential to realize an efficient trigger system to filter events. In conventional data acquisition electronic system, the level-1 trigger that makes the first level of event selection is implemented by hardware and provides trigger signals for readout modules to select event data. As a result, the data rate to be transmitted can be reduced by several orders of magnitude. However, the research on various particle physics experiments shows that the physical event rate and number of electronic channels are increasing. Such as ATLAS that is a particle physical experimental instrument at the Large Hadron Collider at CERN, its number of electronic channels reaches upwards of $10^7$ and raw data rate is up to hundreds GBytes/s [2-3]. In addition, some collider physics experiments, for example GREAT [4] at Jyväskylä and CBM [5] at GSI/FAIR, begin to use a triggerless data collection method called Total Data Readout [6] to solve some problems (e.g. the dead time) caused by the traditional trigger structure. Here, the "triggerless" is not entirely without triggering, but abandoned the hardware trigger, replaced by software trigger in DAQ system. Therefore, the readout scheme meets the challenge of improving readout speed caused by the limited bandwidth of crate backplane.

Advanced Telecommunications Computing Architecture (ATCA) [7] developed by the PCI Industrial Computer Manufacturers Group (PICMG) is a set of industry standard specifications for the next generation of telecommunication network and data center equipment. The ATCA backplane provides point-to-point connections between the boards and supports the dual star, dual-dual star, and mesh topologies. The dual star topology based on 4X InfiniBand links supports 10Gigabit/sec of raw throughput between hub boards and node boards [8]. The ATCA with high performance backplane has been proposed to develop next generation electronics standards for physics [9]. Despite all of these features, however, ATCA has not been widely applied and it popularizes slowly

---
[1] * Supported by National Natural Science Foundation of China (11005107) and independent project of State Key Laboratory of Particle Detection and Electronics (201301)
1) E-mail: xiru@mail.ustc.edu.cn
2) E-mail: cping@ustc.edu.cn

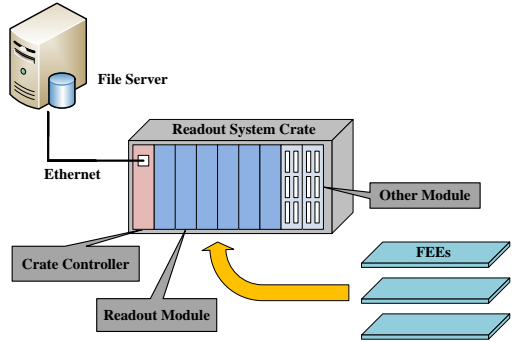

Fig. 1.　Simplified architecture of a typical readout system.

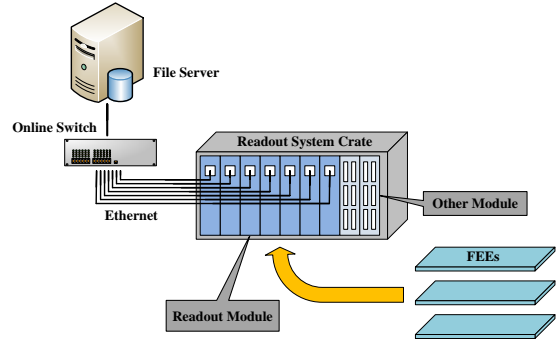

Fig. 2.　Simplified architecture of a new-type readout system.

in particle physics experiments.

The main task of the crate controller in a typical readout system is to gather data from the crate backplane and then transmit them to DAQ system through networks. This is a centralized architecture that needs to improve backplane performance for applications with higher readout speed requirement. Actually, a decentralized architecture can be used for readout system that has advantage for improving readout speed, compared with the centralized one.

In this paper, a high-speed data readout method based on Ethernet is introduced. A simplified readout prototype module is designed and implemented. This module has a 100M Ethernet port on it. Test result shows that the raw data throughput from the readout module to DAQ can reach up to 70Mbps.

## 2 High-speed data readout method

Instead of improving backplane performance, each readout module is designed to have capability of transmitting data to DAQ, as shown in Fig. 2.

To make each readout module supporting network communication, a dedicated CPU is utilized to establish an embedded system. Fig.3 shows the system architecture of the method. It is consisted of two main blocks: embedded CPU and field programmable gate array (FPGA). There are three important transmission channels: a high-speed data transmission channel, a low-speed command transmission channel, and an Ethernet interface respectively. Considering the fact that the throughput of data uploading from FEE to DAQ is much larger that of data downloading to FEE in particle physics experiments. So the high-speed data transmission channel is designed as simplex mode and only used for transmitting data from FPGA to CPU. While the low-speed command transmission channel is also committed as simplex mode for sending command or configuration from

CPU to FPGA. Based on these two dedicated channels, there can be realized a full-duplex data communication mode between CPU and FPGA.

Unrelated with FPGA, the CPU is a standalone chip for maintaining an embedded Linux operation system. The TCP/IP protocol stack is supported in the operation system. This embedded system can be considered as a small computer based on a particular microprocessor core. This chip generally integrates various necessary features and peripherals such as internal memory (e.g., ROM, RAM), input/output control unit, serial bus interface (e.g., serial peripheral interface (SPI), universal asynchronous receiver transceiver (UART), inter-integrated circuit ($I^2C$), and external memory controller etc. The external memory controller is capable of handling several types of external memories and peripheral devices, such as SRAM, Flash, and SDRAM. There are many embedded CPUs can be considered as the candidate for implementing readout module, such as ARM, Power PC, MIPS, Am186/88, and so on.

With the help of embedded Linux system, data can be packaged with TCP/IP protocol and sent to DAQ. But there lies a problem that how to transfer data from FEE to this CPU system efficiently and smoothly in real-time. Considering the simplicity and universality, in this readout architecture, the high-speed data transmission is implemented by using an external SRAM interface which connecting CPU and FPGA together. On one side, FPGA receives data from FEE, re-packages and feeds the data to this SRAM interface. On the other side, CPU receives data from this SRAM interface and relays to DAQ through Ethernet port. The use SRAM interface makes it easier to design or implement readout module, whichever for hardware circuit or FPGA logic. There is no need to use special IP (intellectual property) cores, which degrades the cost or requirement for FPGA. SRAM interface has advantages of simplicity, low cost and high performance as well. In fact, the maximum data throughput of a synchronous SRAM interface can reach up to 1600Mbps with 100MHz running clock and 16-bit data width. On the other side, the maximum data throughput of an asynchronous SRAM interface can be up to 800Mbps with 20-ns read/write cycle period and 16-bit data width.

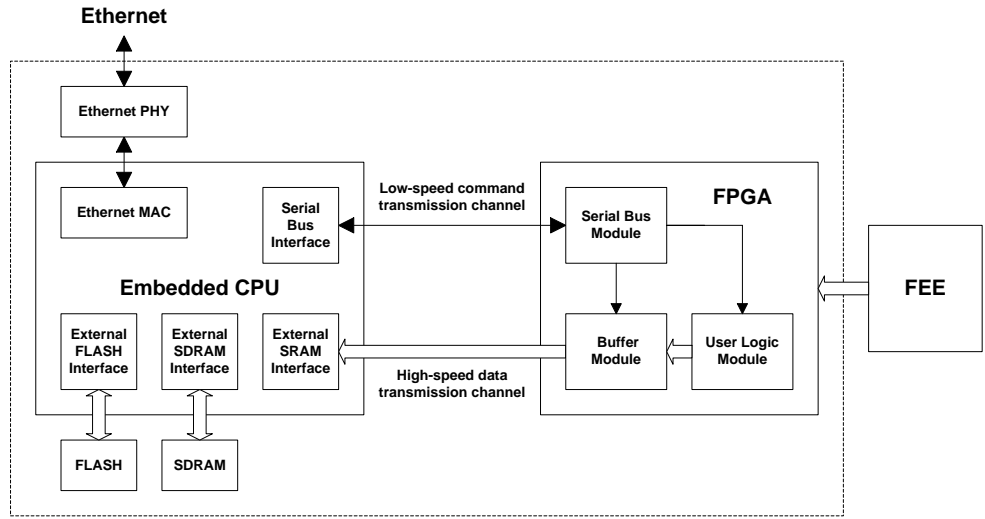

Fig. 3.    System architecture of the high-speed data readout method.

Differing from the data channel, the low-speed command transmission channel is implemented by using a serial data interface between CPU and FPGA. Besides, the CPU system must also support a high-speed Ethernet MAC and physical signal interface, which is used for transmitting/receiving data to/from Ethernet transmission lines.

Nowadays, FPGAs have large resources of logic gates and RAM blocks which make it good for implementing complex digital logic design. In this system architecture, the key task of FPGA is to process data sent from FEE with customized format and protocol, and to send the valid data to CPU through the high-speed data channel for Ethernet transmission.

## 3 Implementation of the module inside FPGA

During transferring, CPU reads data from FPGA via the SRAM interface. However, FPGA isn't a standard SRAM for CPU. It should always be ready and response correctly whenever there is a data receiving or transmitting transaction. A data synchronization mechanism should be guaranteed between CPU and FPGA. Here a hand-shaking protocol is provided. Table 1 lists all of the hand-shaking signals.

Table 1.    Additional hand-shaking signals.

| Signal name | Description |
| --- | --- |
| Read Ready | The read ready signal is used for initiating a data transfer by CPU. |
| IRQ | The interrupt request signal is used for interrupting CPU by FPGA. When FPGA detects that the read ready signal is valid, it will judge whether the data is ready. If it is ready, FPGA makes the interrupt request signal valid. Then, when FPGA waits for the read ready signal is invalid, it will make the interrupt request signal invalid. |
| CLK | The clock signal is used for synchronizing signal timing of the SRAM interface. If the SRAM interface is an asynchronous SRAM interface, CPU needs to provide a clock signal for FPGA. |

In FPGA, the main logic module is called buffer module. It is in charge of caching data from the user logic module and sending them to CPU through the SRAM bus. Fig.4 shows the detailed structure of this buffer module. It consists of three parts: FIFO, FIFO controller, and state machine. FIFO is used for caching data. It is designed as an asynchronous FIFO. It provides a general FIFO interface (including signals of datain, wrclk, wrreq, wrfull) to the user logic module from the input side. While from the output side, there is a data bus connecting it with the SRAM interface. To guarantee operate synchronously, the clock of the FIFO output side is provided by CPU.

FIFO controller generates the read request signal for FIFO according to the control signal timing of the SRAM bus, and doesn't need to use the address signal of the SRAM bus. When the read ready signal is valid, the state machine will judge whether the data in FIFO is ready through comparing the transmission length with the FIFO count. Then the state machine makes IRQ signal valid to respond the read data request of CPU. In addition, the transmission length can be set to a default value or configured by the command.

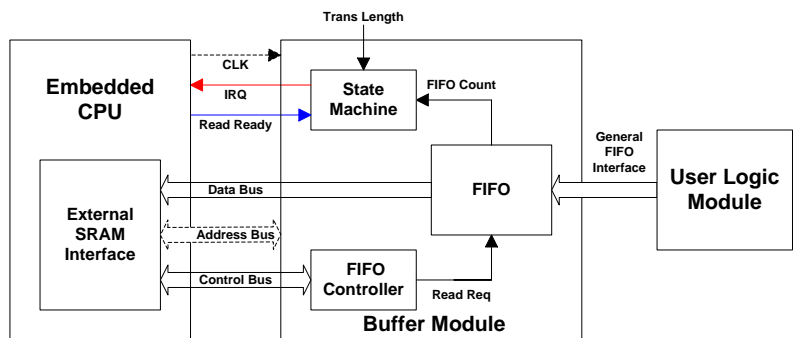

Fig. 4.　Block diagram of the buffer module.

# 4 Implementation of network transmission

An embedded Linux operating system will be installed on this microcomputer system. Linux OS provides a complete, powerful network functions for users and its performance in real-time applications is improving with the development of new scheduling algorithm. The Linux kernel is released under an open source license, so anyone can read and modify its code.

## 4.1 Design of device-driver

A device driver (commonly called driver) provides a software interface to hardware devices and tells the operating system and other software how to access hardware without needing to know precise details of the hardware being used. In Linux environments, drivers are built as modules that can be easily loaded in or unloaded from the kernel during running time.

For the Linux OS, the buffer module can be viewed as a device and not use the generic SRAM driver. Therefore a special driver needs to be prepared for controlling and managing the buffer module. The driver can be classified as a character driver. Fig.5 shows the structure of the driver in kernel space.

There are several major modules in this structure: initialization module, exit module, open module, release module, read module, memory mapping (mmap) module, and interrupt handler.

First, the initialization module has to configure the device resources, for example, set IRQ pin, set read ready pin, configure SRAM interface. Then the module allocates a major and a minor number for this char device. Finally, the initialization is finished after registration of the char device.

Once the module has been installed successfully, the driver is ready for work. The user software in user space uses the open method to request the I/O memory resources according to the memory mapping of CPU. Then the open method completes an important process: requesting an interrupt number from the system and installing an interrupt handler. Finally, if the open method is called successfully, it will return a file descriptor that establishes an access path to the device.

In our design scheme, the user software doesn't use the read method to acquire data, but accesses the address pointer returned from the mmap method to fetch data. Data obtaining procedure as follows: first, the user software tries to read the device and wait for its return. On the kernel side, the read module first set read ready pin valid to inform FPGA that it is ready to receive data. Next it uses a macro called *wait_event_interruptible* to put the process into an interruptible sleep. Once the data in the buffer module is ready, an interrupt is generated and the corresponding handler is immediately evoked. Then the interrupt handler sets read ready pin invalid and wakes the read module up by the function called *wake_up_interruptible*. Finally the read module returns a status flag to the user software. At this point, the read method is complete, and then the user software can access the address pointer to read data.

## 4.2 Design of software

One of the main tasks of the readout module designed by the high-speed data readout method is to transfer data to the concentration center or PC farms via Ethernet. To maximize transfer performance, the software residing in the embedded Linux system must be designed in a simple and parallel structure, as shown in Fig.6.

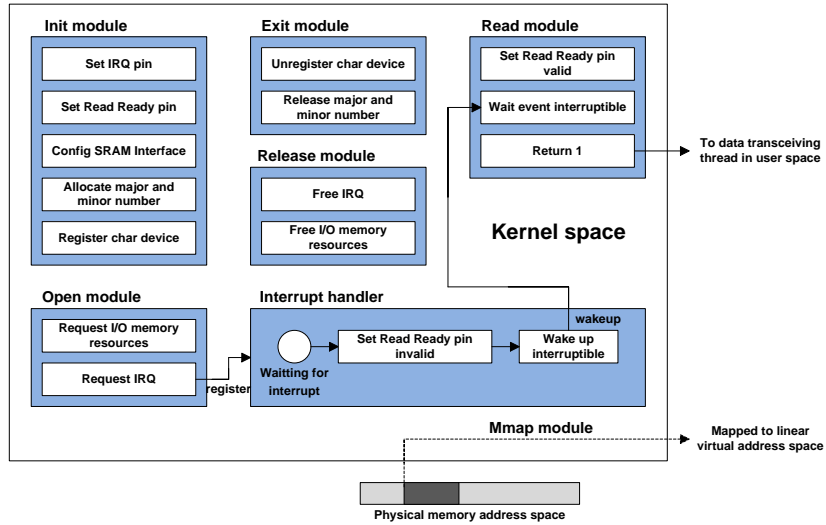

Fig. 5.   Driver in kernel space.

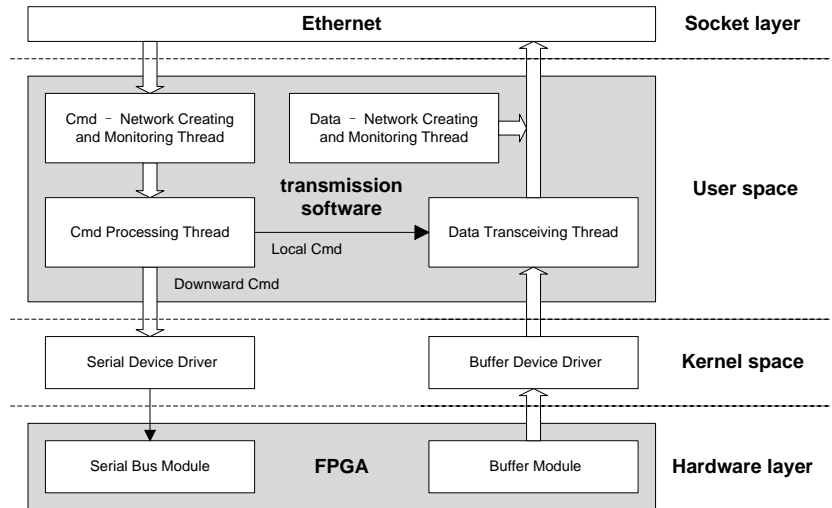

Fig. 6.   Structure of the transmission software in user space.

There are 4 kinds of threads in this transmission software: command–network creating and monitoring thread, command processing thread, data–network creating and monitoring thread, and data transceiving thread. When the transmission software start running, these threads will automatically be created with detached attribute and are run in parallel in the user space. Command–network creating and monitoring thread creates a server socket and then builds a command channel through accepting connections from the PC client. This command channel is only responsible for receiving commands. After the channel is built, the thread monitors the channel on and off. Once the channel is disconnected, the thread will close the original channel and rebuild a new channel. Data–network creating and monitoring thread also completes similar work.

The task of command processing thread is to process commands. If the command is a local command used for configuring and controlling the software, the thread will analyze the command and then make an appropriate response. If the command is a downward command used for configuring and controlling the hardware, the thread will send the command to FPGA via the write method related to the serial device.

The task of data transceiving thread is to receive data from FPGA and then transmit them to PC via Ethernet. During the initialization of the thread, it creates a file descriptor related to the buffer module via using the open method and gets an address pointer related to physical address space of the buffer module via using the mmap method. Then in a loop, it first needs to ensure that the data in the buffer module is ready through the read method using the file descriptor, and next transfer this piece of data to PC through the write method using the data server socket and address pointer, and then return to the beginning of the loop.

# 5 Experiments and evaluation

For the purpose of evaluating this high-speed readout method, a simplified readout module is designed and implemented with an AT91RM9200 and EP3C40F780C8 chip.

The AT91RM9200 chip [10] is a microcontroller introduced by Atmel Corporation, which is based on the ARM920T 32-bit RISC processor with 16KB instruction and 16KB data cache memories and memory management unit (MMU). Its speed can achieve 200 MIPS when working at 180 MHz frequency. The chip contains an external bus interface (EBI) dedicated to interfacing external memory devices, four programmable external clock signals, seven external interrupt sources, four 32-bit I/O controllers, a 10/100 Base-T Ethernet MAC, four universal synchronous/asynchronous receiver transceivers (USART), and so on. The EBI integrates three external memory controllers: static memory controller (SMC, supported 16-bit asynchronous SRAM), SDRAM controller, and burst flash controller. Furthermore, the EBI handled data transfers with up to eight external devices through eight chip select lines, each assigned to eight address spaces defined by the embedded memory controller. Therefore, this chip is able to satisfy the application requirements of the readout method.

The EP3C40F780C8 chip [11] is a FPGA of Cyclone III device family of Altera Corporation. The Cyclone III device family is a high functionality, low-power and low-cost FPGA family, which is based on Taiwan Semiconductor Manufacturing Company (TSMC) 65-nm low-power (LP) process technology. The chip contains a great variety of logical unites (nearly 40,000 LEs), 126 M9K memory blocks (9 Kbits), 126 $18\times18$-bit multiplier units, 4 PLLs, 20 global clock networks, and 780 external lead feet.

Here, the high-speed data transmission channel is implemented by using the static memory controller, an external interrupt source, a programmable external clock signal and a programmable I/O line of the AT91RM9200 chip. Moreover, the low-speed command transmission channel is implemented by using a USART of the AT91RM9200 chip. Fig.7 shows the prototype of the readout module.

To measure the network transfer performance of the readout module, the Ethernet port of the readout module is directly connected to a PC using a good wiring. Then an embedded Linux OS with kernel version 2.6 is running on the readout module. As shown in Fig.8, route 1 denotes the throughput of SRAM bus from EP3C40F780C8 to AT91RM9200, route 2 denotes the throughput of 100M Ethernet from AT91RM9200 to PC, and route 3 denotes the throughput of 100M Ethernet from EP3C40F780C8 to PC. Moreover, the data in the buffer module is always ready.

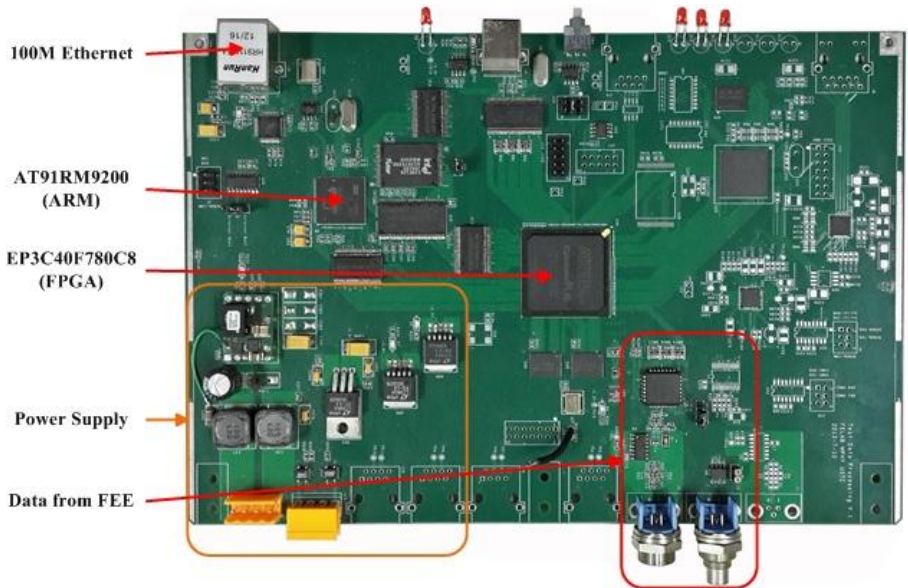

Fig. 7.    Prototype PCB of the readout module.

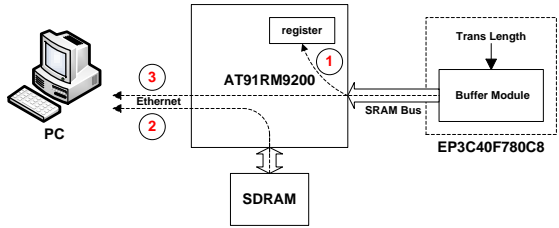

Fig. 8.    Measurement scheme.

Figure 9 and 10 show the measurement result of the route 1, the route 2, and the route 3 limited in a wide range of transmission lengths. There are two main time costs in the route 1 and route 3: data synchronization time and data transmission time. When the transmission length is relatively small, such as 256 bytes, a large proportion of time is spent in data synchronization. With increasing the transmission length, the proportion of time spent in data synchronization is smaller and smaller, at the same time, the throughput of the route 1 and route 3 is bigger and bigger and finally approaches a stable value. For transmission lengths larger than 16384 bytes, the throughput of the route 2 and route 3 can reaches up to 70Mbps.

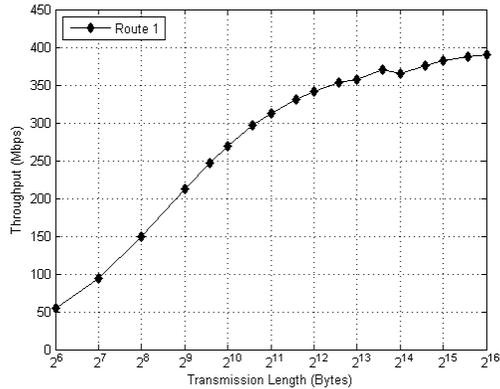

Fig. 9.    Data throughput of SRAM interface.

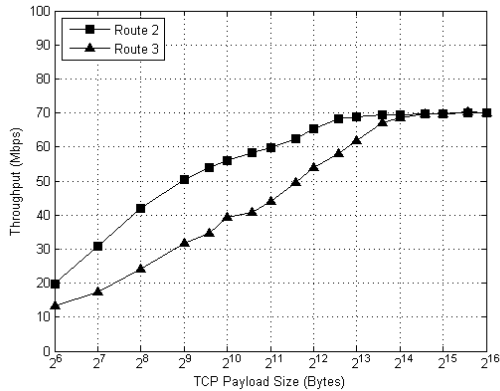

Fig. 10.    Network throughput.

# 6 Conclusions and future work

A high-speed data readout method based on embedded CPU and FPGA is presented in this paper. This method makes each readout module being capable of communicating with DAQ system through networks. It has advantages of simplicity, universality, expansibility and low cost. It is suitable for various applications of data readout in particle physics experiments. To verify and evaluate this method, a prototype readout module is designed and implemented. Test results show that this module can support up to 70Mbps valid data throughput from the readout module to DAQ.

To further improve the data throughput (e.g. 1000Mbps), a more powerful CPU should be adopted for data and protocol processing. To achieve higher performance of transmitting data to CPU from FPGA with this presented method, there's little modification except for improving the clock frequency and modifying logical behavior of the SRAM interface according to CPU.